\documentclass[aps,prb,twocolumn,showpacs,preprintnumbers,superscriptaddress]{revtex4}

\usepackage{graphicx}
\usepackage{dcolumn}
\usepackage{bm}

\begin{document}


\title{Effect of pressure on the steplike magnetostriction  of  single crystalline bilayered manganite(La$_{0.4}$Pr$_{0.6}$)$_{1.2}$Sr$_{1.8}$Mn$_{2}$O$_{7}$  }

\author{Y.Yamato}
\author{M.Matsukawa} 
\email{matsukawa@iwate-u.ac.jp }

\author{T.Kumagai}
\affiliation{Department of Materials Science and Engineering, Iwate University , Morioka 020-8551 , Japan }
\author{R.Suryanarayanan}
\affiliation{Laboratoire de Physico-Chimie de L'Etat Solide,CNRS,UMR8182
 Universite Paris-Sud, 91405 Orsay,France}
\author{S.Nimori}
\affiliation{National Institute for Materials Science, Tsukuba 305-0047 ,Japan} 
\author{M.Apostu}
\affiliation{Department of Physical, Theoretical and Materials Chemistry, Faculty
of Chemistry, Al. I. Cuza University, Carol I, 700506 Iasi, Romania
}
\author{A.Revcolevschi}
\affiliation{Laboratoire de Physico-Chimie de L'Etat Solide,CNRS,UMR8182
 Universite Paris-Sud, 91405 Orsay,France}
\author{K. Koyama}
\author{ N. Kobayashi}
\affiliation{Institute for Materials Research, Tohoku University, Sendai  
980-8577, Japan}

\date{\today}

\begin{abstract}
We report  the effect of pressure on   the steplike  magnetostriction  of single crystalline bilayered manganite(La$_{0.4}$ Pr$_{0.6}$)$_{1.2}$Sr$_{1.8}$Mn$_{2}$O$_{7}$ , for our understandings of the ultrasharp nature of the field-induced first-order transition from a paramagnetic insulator to a ferromagnetic metal phase.
The application of pressure suppresses a steplike transformation and causes a broad change in the magnetostriction.  The injection of an electric current to the crystal also weakens the steplike variation in both the magnetostriction and magnetoresistance. 
The stabilization of ferromagnetic interaction or the delocalization of charge carriers is promoted with the applied pressure or applied current, resulting in the suppressed 
steplike behavior.  Our findings suggest that the step phenomenon is closely related to the existence of localized carriers such as the short-range charge-orbital ordered clusters. 
\end{abstract}

\pacs{}
\maketitle
\section{ INTRODUCTION}


Perovskite manganites show a great variety of fascinating  properties such as colossal magnetoresistance (CMR) effect and charge-ordered insulating  phase \cite{TO00}. The most interesting one is the existence of a phase-separated state, the coexistence of antiferromagnetic charge-orbital ordered (COO) insulating   and  ferromagnetic (FM)  metal regions\cite{DA01}. Several recent studies on metamagnetic transition in phase-separated manganites have revealed that  ultrasharp steps in magnetization curves appear at low temperatures \cite{MAH02,HA03,FI04,GHI04,SU06,LI06}.
To account for  this, a martensitic model due to local strain fields stored in the lattice between competing COO and FM phases with their different unit cells has been proposed though questions have been raised against this model.

For our understanding of the dynamics of a steplike first-order transition from a paramagnetic insulating (PMI) to a ferromagnetic metal(FMM) phase in CMR manganites, we examine the pressure effect on  $c$ axis magnetostriction of
 single crystalline (La$_{0.4}$Pr$_{0.6}$)$_{1.2}$Sr$_{1.8}$Mn$_{2}$O$_{7}$.  
%


For the Pr-substituted (La$_{0.4}$Pr$_{0.6}$)$_{1.2}$Sr$_{1.8}$Mn$_{2}$O$_{7}$ crystal,  a spontaneous ferromagnetic metal phase (originally present with no Pr substitution) disappears at  ground state but a field-induced PMI to FMM transformation is observed over a wide range of temperatures\cite{MO97,AP01}.
A magnetic  ($H,T$) phase diagram, established from  magnetic measurements, is separated into three regions labeled as PMI, FMM , and bistable states, as shown in Fig.1 of ref.\cite{MA05}. The proximity of free energies between the PMI and FMM states is of importance to realize the bistable state where the FMM and PMI states coexist. 
A stronger field is needed to induce the PMI to FMM transition at low temperatures since the thermal energy is reduced upon lowering temperature.  
\section{EXPERIMENT}
Single crystals of (La$_{1-z}$,Pr$_{z}$)$_{1.2}$Sr$_{1.8}$Mn$_{2}$O$_{7}$ ($z$=0.6) were grown by the floating zone method using  a mirror furnace. 
The calculated lattice parameters of the tetragonal crystal structure of the crystals used here were shown in a previous report\cite{AP01}. 
 The dimensions of  the $z$=0.6 sample are 3.4$\times$3 mm$^2$ in the $ab$-plane and 1mm along the $c$-axis.  Measurements of magnetostriction along the $c$-axis  were done by means of a conventional strain gauge method at the Tsukuba Magnet Laboratory, 
the National Institute for Materials Science (NIMS) and at the High Field Laboratory for Superconducting Materials, Institute for Materials Research, 
Tohoku University.
 Except for an initial cooled down, the sample was zero-field cooled from 200K down to low temperatures and   we then started  measuring the isothermal
magnetostriction upon increasing (or decreasing) the  applied fields, parallel to the $c$-axis. 
Next, warming the sample from low temperatures up to 200K in the absence of field and keeping the sample fixed at 200K 
for  2 hours,the sample was then  cooled down to the low T. A typical cooled time is needed for about 2 hours.
Checking the stability of sample's temperature,  we restart measuring the magnetostriction. For each measurement, we repeated the same procedure.
The normal sweep rate was set to be 0.26 T/min. 
Hydrostatic pressures  in  the magnetostriction  experiment were applied  by a clamp-type cell using Fluorinert as a pressure transmitting medium. The pressure was calibrated by the critical temperature of lead. 
After the magnetostriction measurement under the pressure, we restarted measuring $dL_{c}(H)$ using the same clamp-type cell but without the applied pressure. Next, we examined the $c$-axis magnetostriction as a function of the applied current in the $ab$ plane at ambient pressure. Once the sample was removed from  the clamp-type cell, it was set in a conventional cryostat for magneto transport measurements. 
On both ends of the sample the electrodes for injection of the electric current  were  formed  using a gold paste and Cu  wires were then attached with silver paste. 
For magnetoresistance measurements, we carried out the same temperature profile as the magnetostriction experiments. 
\begin{figure}[ht]
\includegraphics[width=8cm,clip]{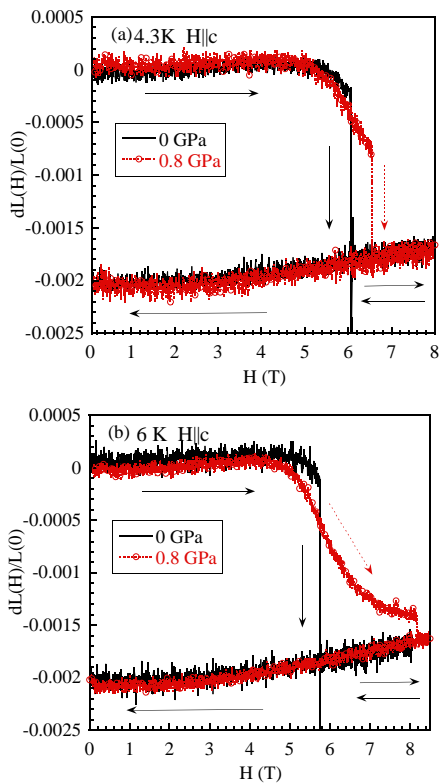}%
\caption{(color online) the $c$-axis magnetostriction, $dL_{c}(H)/L_{c}(0)$, of (La$_{0.4}$Pr$_{0.6}$)$_{1.2}$Sr$_{1.8}$Mn$_{2}$O$_{7}$,  both under ambient pressure and a hydrostatic pressure of 0.8 GPa at low temperatures (a) 4K and (b) 6K. The  applied field is parallel to the $c$-axis ($H||c$) .
}
\label{Hc}
\end{figure}%

\section{Results and discussion}
Figure \ref{Hc} shows the $c$-axis magnetostriction, $dL_{c}(H)/L_{c}(0)$, of (La$_{0.4}$Pr$_{0.6}$)$_{1.2}$Sr$_{1.8}$Mn$_{2}$O$_{7}$,  both under ambient pressure and a hydrostatic pressure of 0.8 GPa at lower temperatures, where the  applied field is parallel to the $c$-axis ($H||c$). Here, the value of $dL_{i}(H)$ is defined as $L_{i}(H)-L_{i}(0)$. 
First of all, the steplike lattice transformations appear at 4 K and 6 K at ambient pressure, as shown in Fig.\ref{Hc}.

\begin{figure}[ht]
\includegraphics[width=8cm]{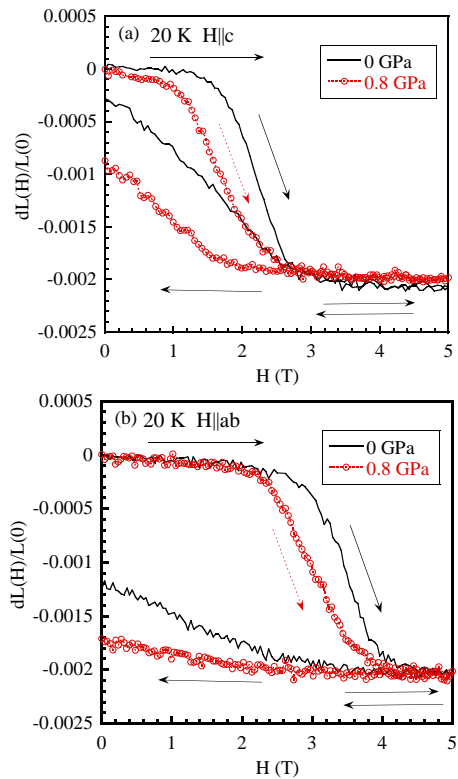}%
\caption{(color online) the $c$-axis magnetostriction, $dL_{c}(H)/L_{c}(0)$, of (La$_{0.4}$Pr$_{0.6}$)$_{1.2}$Sr$_{1.8}$Mn$_{2}$O$_{7}$,  both under ambient pressure and a hydrostatic pressure of 0.8 GPa at 20 K. (a)$H||c$ and (b)$H||ab$.
}
\label{HT}
\end{figure}

 On the other hand, the application of external pressure on the crystal substantially suppresses the steplike transition,
causing a broad variation in the magnetostriction. At 4.3 K, the critical field, $H_{c}$, is increased from 6.1 T at ambient pressure up to 6.6 T at 0.8 GPa.
At 6 K, a huge steplike transition vanishes under the pressure and a continuously smooth variation is observed at high fields, followed by the appearance of a tiny step around 8 T. 
However, the characteristic field signifying the onset of the field-induced metamagnetic transition, which is not the critical field of the step transitions,  is rather lowered at the presence of the applied pressure, as the pressure data measured at high-$T$  in Fig.\ref{HT}. 

\begin{figure}[ht]
\includegraphics[width=9cm,clip]{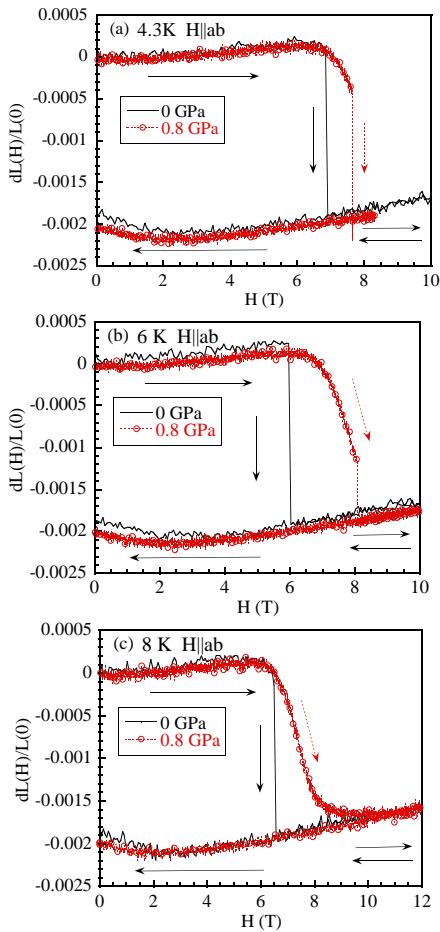}%
\caption{(color online) the $c$-axis magnetostriction, $dL_{c}(H)/L_{c}(0)$, of (La$_{0.4}$Pr$_{0.6}$)$_{1.2}$Sr$_{1.8}$Mn$_{2}$O$_{7}$,  both under ambient pressure and a hydrostatic pressure of 0.8 GPa at low temperatures (a) 4K (b) 6K and (c) 8K. The  applied field is parallel to the $ab$-plane ($H||ab$) .
}
\label{Hab}
\end{figure}%
 When the magnetic field is applied to the $ab$ plane ($H||ab$), we obtain similar results, as shown in Fig.\ref{Hab}.  These findings are indicative of the occurrence of steplike transformation almost independent of the easy axis of magnetization.  The small differences of the higher critical fields in the case of $H||ab$ probably arises from the easy axis of $M$ lying along the $c$-axis.  Under the application of 0.8 GPa, we obtain both $H_{c}\sim$7.6 T at 4.3 K and 8.1 T at 6 K accompanied by a degradation of a discontinuous jump, and the disappearance of any steplike profile in $dL_{c}(H)/L_{c}(0)$ curve at 8 K.

\begin{figure}[ht]
\includegraphics[width=9cm]{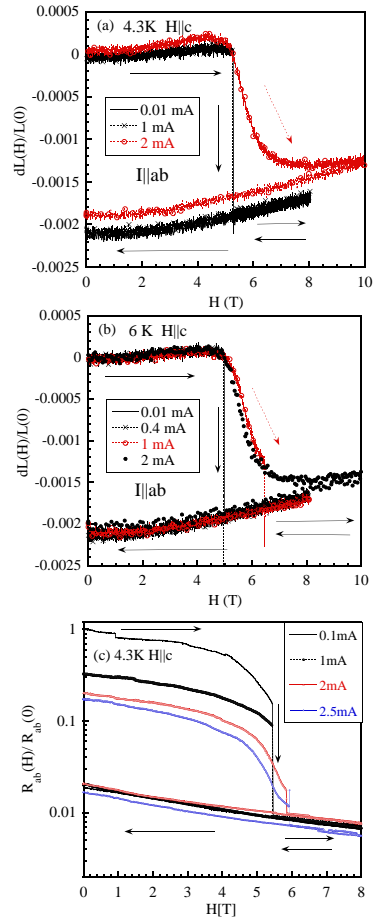}%
\caption{(color online) the $c$-axis magnetostriction, $dL_{c}(H)/L_{c}(0)$, of (La$_{0.4}$Pr$_{0.6}$)$_{1.2}$Sr$_{1.8}$Mn$_{2}$O$_{7}$,  under the electric current applied in the $ab$ plane  at low temperatures (a) 4K and (b) 6K. The  applied field is parallel to the $c$-axis. (c) the $ab$ plane magnetoresistance $R_{ab}$ of the $z=0.6$ sample at 4.3K as a function of the applied current from 0.1 up to 2.5 mA. The $R_{ab}$(H) data are normalized by $R_{ab}$(0), which is the zero-field resistance at the lowest applied current (0.1 mA). ($H||c$)}
\label{dLI2}
\end{figure}%

In our previous paper\cite{MA07}, we have found out that the application of pressure enhances a field-induced ferromagnetic state of Pr-substituted La$_{1.2}$Sr$_{1.8}$Mn$_{2}$O$_{7}$. Our findings are understood from the viewpoint that the double-exchange interaction-driven FM state is strengthened by the applied pressure.
 For the magnetic field cooled sample, the remarkable step observed in $dL_{c}(H)/L_{c}(0)$ is monotonically decreased upon increasing the cooling field from 1 T up to 1.6 T, accompanied by a nonlinear dependence of  $H_{c}$\cite{MA07L}. 
The steplike behavior vanishes when the cooling field exceeds 1.6 T. For each field-cooled run, we expect that  the FM region is initially formed within the PMI matrix before the magnetostriction measurements are carried out.  A larger volume fraction of the FM phase causes a more suppressed variation in $dL$. Thus, a pressure-induced suppression in the ultrasharp magnetostriction is a reasonable result since the applied pressure enhances the FM state.
 
Next, we carry out  the magnetostriction measurement under the applied electric current, to examine a relationship between charge transport and steplike lattice deformation appeared in CMR manganites. In the case of ambient pressure, the $c$ axis magnetostriction data at 4.3K are presented  in Fig.\ref{dLI2}(a) as a function of excited current ($I$=0.01, 1 and 2 mA).
Here,  the electric current is applied in the $ab$ plane parallel to the MnO$_{2}$ double layers. 
Injection of the electric current over 2 mA exhibits no discontinuous profile in $dL_{c}(H)/L_{c}(0)$ .  
We note that the bath temperature is very stable against the applied current up to 2 mA within $\sim0.5\%$ and Joule heating of the sample affects no contribution to the present behavior. 

For comparison, let us now display in Fig.\ref{dLI2}(c) the $ab$ plane magnetoresistance $R_{ab}$ of the $z=0.6$ sample as a function of the applied current from 0.1 up to 2.5 mA. 
First of all,  the application of higher current gives rise to a degradation of  the magnetoresistive steps observed at lower currents, which is almost consistent with the magnetostriction data shown in Fig.\ref{dLI2}(a). Next, the zero-field resistance measured at 2 mA  shows a substantial drop by $\sim 80\%$, in comparison to $R_{ab}$(0) at the lowest current (0.1 mA). This result  indicates the current-induced destabilization of PMI matrix although the physical role of the application on an electric field is not  clear. Similar results are reported in a systematic study on nonlinear transport of a non charge-ordered manganite Pr$_{0.8}$Ca$_{0.2}$MnO$_{3}$, which is not so far from the CO region\cite{ME02}.
In a lightly Pr-substituted crystal (La$_{0.8}$Pr$_{0.2}$)$_{1.2}$Sr$_{1.8}$Mn$_{2}$O$_{7}$ with $T_{c}\sim90$K, we also have obtained a current-induced remarkable drop in the $ab$ plane resistance at selected temperatures just above $T_{c}$.
(not shown here)
The formation of short-range charge ordering in the paramagnetic phase of an optimally doped parent crystal La$_{1.2}$Sr$_{1.8}$Mn$_{2}$O$_{7}$ has been observed in a previous study\cite{VA99}.
Accordingly, we expect that the application of the electric current removes the localizes carriers, such as the COO insulating clusters, from the PMI phase of the Pr-substituted crystal, causing a substantial reduction in the zero-field resistance at higher currents.

Furthermore, neutron scattering measurements on a bilayered manganite La$_{2-2x}$Sr$_{1+2x}$Mn$_{2}$O$_{7}$ ($x=0.38$) have revealed that the CE-type COO clusters freeze upon decreasing temperature from  $T\sim310$K, preventing the formation of a long-range COO state \cite{AR02}. The authors have also argued that  the PMI state of the crystal without Pr substitution  is caused by an orbital frustration and stabilized down to the FMM transition temperature $T_{c}\sim114$K. 
For the present Pr-substituted sample, one believes that the short-range COO clusters accompanied by the orbital frustration are present within the PMI matrix.  As discussed in ref. \cite{MA07L}, we suppose that the orbital frustration has some relationship with the critical instability of the metastable state of the free energy,resulting in the steplike phase transition. 
The application of hydrostatic pressure stabilizes the ferromagnetic interaction within the paramagnetic matrix and the applied current reinforces an itinerant state of charge carriers within the insulating matrix. In other words, these external variables probably lead to a collapse  of the CE-type antiferromagnetic charge-ordered clusters if these type clusters are distributed  within the sample.
In addition, Mahendiran et al. have claimed that  
the charge-orbital ordered domains in the Co-doped Pr$_{0.5}$Ca$_{0.5}$MnO$_{3}$ are smaller than the undoped Pr$_{0.7}$Ca$_{0.3}$MnO$_{3}$, suggesting that the onset temperature of the step transitions in the Co-doped compound is higher than that 
 in the undoped one \cite{MAH02}. 
We emphasize that the distribution of short-range COO clusters embedded in the matrix provides  a significant clue of our understandings of the steplike transformations in CMR manganites. 
The ultrasharp  PMI-FMM transition induced by the magnetic field  at low temperatures is in its origin quite different from the standard IM transition associated with the metamagnetic transition at higher $T$.

In summary,we have demonstrated  the influence of pressure  on   the steplike  lattice deformation   of single crystalline bilayered manganite(La$_{0.4}$ Pr$_{0.6}$)$_{1.2}$Sr$_{1.8}$Mn$_{2}$O$_{7}$ .
The external perturbations such as the application of pressure and electric current give a substantial suppression on the steplike transition, resulting in a broad change in the magnetostriction.  
It is a future problem to resolve what role the CO clusters within the PMI phase play in the occurrence of the steplike phenomenon observed.

This work was supported by a Grant-in-Aid for Scientific Research from Japan Society of the Promotion of Science.

\end{document}